\documentclass[article]{aa}
\usepackage{graphicx,txfonts,natbib,color}
\usepackage{booktabs}

\begin{document}

\title{Effects of stellar flybys on planetary systems: 3D modeling of the
  circumstellar disks damping effects.}

\author{G. Picogna\inst{1}, F. Marzari\inst{1}}

\institute{Dipartimento di Fisica, University of Padova, Via Marzolo 8,
  35131 Padova, Italy}

\titlerunning{Circumstellar disks and stellar flybys.}
\authorrunning{G. Picogna and F. Marzari}

\abstract 
 {Stellar flybys in star clusters are suspected to affect the orbital
  architecture of planetary systems causing eccentricity excitation and
  orbital misalignment between the planet orbit and the equatorial plane 
  of the star.}
 {We explore whether the impulsive changes in the orbital elements of
  planets, caused by an hyperbolic stellar flyby, can be fully damped by 
  the circumstellar disk surrounding the star. 
  The time required to disperse stellar clusters is in fact comparable to
  circumstellar disk's lifetime.
  Since we perform 3D simulations, we can also test the inclination 
  excitation and damping.}
  {We have modelled in 3D with the SPH code VINE 
  a system made of a solar type star surrounded by
  a low density disk with a giant planet embedded in it
  approached on a hyperbolic encounter trajectory by a 
  second star, of similar mass and with its own disk.
  Different inclinations between the disks, planet orbit and star 
  trajectory have been considered to explore various
  encounter geometries. 
  We focus on an extreme configuration where a very deep stellar flyby 
  perturbs
  a Jovian planet on an external orbit. This allows
  to test in full the ability of the disk 
  to erase the effects of the stellar encounter.}
  {We find that the amount of mass lost by the disk during the stellar 
  flyby is less than in 2D models where a single disk was considered.
  This is mostly related to the mass exchange between the two disks at the
  encounter.
  The damping in eccentricity is slightly faster than in 2D models and it
  occurs on timescales of the order of a few kyr.
  During the flyby both the disks are warped due to the mutual interaction
  and to the stellar gravitational perturbations, but they quickly relax 
  to a new orbital plane.
  The planet is quickly dragged back within the disk by the tidal
  interaction with the gas.
  The only trace of the flyby left in the planet system, after about 
  $10^4$ yr, is only a small misalignment, lower than $9^o$,
   between the star equatorial
  plane and the planet orbit. }
  {In a realistic model based on 3D simulations of star--planet--disk 
  interactions, we find that stellar flybys cannot excite significant 
  eccentricities and inclinations of planets in stellar clusters. 
  The circumstellar disks hosting the planets damp on a short timescale
  all the step changes in the two orbital parameters produced during any 
  stellar encounter.
  All records of past encounters are erased.}


\maketitle

\section{Introduction} 
\label{intro}

  Stars generally do not form alone but in bound clusters on a timescale
  of about $1$ Myr \citep{hille97, palla00}. 
  The embedded cluster phase lasts only a few Myr and, according to 
  \cite{allen07}, \cite{fall09}, \cite{chan10}, and \cite{duk12}, after
  about $10$ Myr, $90\%$ of stars born in clusters have dispersed into the
  field.
  This occurs because, after a few crossing times \citep{allen07}, 
  embedded clusters lose their gaseous component, the system becomes 
  super--virial and the cluster rapidly dissolves. 
  A cluster, before its dissipation,  is a crowded place and its dynamical
  evolution is characterized by frequent stellar flybys.
  The close encounters can have significant effects on both disks and 
  planetary systems surrounding the stars.
  They can alter the disk morphology causing also mass loss and excite
  planet orbits.
  It is expected that, in the period of higher frequency of stellar 
  encounters, circumstellar disks are still present around stars.
  The lifetime of disks is in fact comparable to the cluster bound phase  
  being in the range $3$--$6$ Myr, as suggested by \cite{hll06}, or even
  longer, up to $10$--$12$ Myr, as recently argued by \cite{bell13}.
  Therefore, to explore the effects of close stellar encounters on
  planetary systems in clusters, one has to account for the presence of
  circumstellar disks around the stars. 

  Recent N--body numerical modeling of the dynamical evolution of  
  planetary systems around stars in clusters 
  \citep{malm1,malm2,malm3,zaktre} have shown that scattering interactions
  with other stars  may excite the eccentricity and inclination of planets
  populating the outer parts of the system. 
  The steep changes in the orbital elements would be observable also at
  present and might contribute to explain the large eccentricities of
  extrasolar planets discovered so far and the presence of misaligned     
  systems. 
  However, \cite{marpi13} have shown that the planet's eccentricity,  
  excited during a close stellar flyby, is damped on a short timescale  
  ($\sim 10$ kyr) by the interaction with the circumstellar disk 
  harbouring the planet in spite of the disk's low density related to its 
  viscous evolution and photoevaporation occurred during the planet 
  growth. 
  As a consequence, the coincidence of the cluster lifetime with that of
  the circumstellar disks plays against a significant role of stellar
  flybys in affecting the present planet dynamical architecture. 

  The simulations presented in \cite{marpi13} were based on the 2D code
  FARGO and considered a single disk around the star with planets while
  the perturber star was assumed diskless. In this paper we extend
  that model using a 3D SPH code and including two disks, one around each
  star.
  This allows us to better model close stellar flybys where the two disks
  closely interact.
  In addition, we can also explore the excitation of the planet
  inclination and its subsequent damping due to the interaction with the
  disk \citep{marne09,klebi11,cres07}.
  We concentrate in this paper on the effects of a single deep close 
  stellar encounter on a planet still embedded in its birth disk.
  If the effects of an extremely close flyby can be erased by the disk, we
  expect that also the cumulative effects of multiple but more distant
  encounters can be absorbed by the disk.
  We will also explore the effects
  on the disk morphology of the stellar tidal force and of the 
  mutual disk interaction during the encounter. 

  The paper is organized as follows.
  In Sect. 2 we describe the numerical algorithm used to model the disks 
  and planet evolution during and after a hyperbolic close encounter with
  another star. 
  In Sect. 3 we outline the different flyby configurations we have modeled
  while Sect 4,5 and 6 are devoted to the presentation of the numerical
  outcomes.
  In Sect. 7 we discuss our results and their implications for the
  evolution of planetary systems in stellar clusters. 

\section{The numerical model} 
\label{model}

  The hydrodynamical code used to model the evolution of the disks 
  surrounding both stars is VINE \citep{Wetzstein09, Nelson09}.
  It is based on the smoothed particle hydrodynamics (SPH) algorithm,
  which solves the hydrodynamical equations by replacing the fluid with
  a set of particles \citep{Gingold77}.
  The code has been updated to improve momentum and energy conservation
  as described in \cite{pima13}.
  Variable smoothing and softening lengths have been introduced as in 
  \cite{Price04}, \cite{Springel02} and \cite{Price07}.
  To realistically model the gas accretion onto the planet, we 
  implemented the algorithm described in \cite{aylba09}.
  The planet potential is modified adding a ``surface'' potential term in 
  the following way:
    \begin{equation}\label{pot}
      F_{\rm r} = -\frac{G M_{\rm p}}{r^2}\left[1 - 
        \left(\frac{2 R_{\rm p} -r}{R_{\rm p}}\right)^4 \right]
    \end{equation}
  for $r < 2 R_{\rm p}$.
  $M_{\rm p}$ and $R_{\rm p}$ are the initial mass and radius of the 
  planet, respectively.
  When an SPH particle reaches the surface of the planet ($R_{\rm p}$) the 
  net force is $0$, whereas if $r$ becomes smaller than $R_{\rm p}$, the 
  potential becomes strongly repulsive and the particle is driven back to 
  the planet surface.  

  The disk around each star has an initial density at $1$ AU of
  $\rho_0 \sim 1 \times 10^{-11}$ g/cm$^3$, approximately $1/100$ of the 
  Minimum Mass Solar Nebula.
  Its radius extends out to $40$ AU with a density profile declining as
  $\rho(r) = \rho_0 r^{-3/2}$, corresponding to a superficial density 
  decreasing as $\Sigma = \Sigma_0 r^{-1/2}$. 
  The total mass of each disk is $7 \times 10^{-3}$ $M_{\odot}$, about 
  $14$ times less massive than the disks modeled in \cite{forgan09}.
  The reason of such a low density is justified by the presence of a 
  fully formed giant planet in the disk. 
  In our scenario, core--accretion and gas infall already occurred and
  most of the gas is expected to have been either accreted by the star 
  because of viscous evolution or dispersed by photo--evaporation. 
  In addition, if eccentricity and inclination damping occurs in light 
  disks, it would be even more efficient in massive disks. 
  In modeling these light disks we can neglect the effects of 
  self--gravity and adopt the isothermal approximation since they are 
  optically thin. 
  The disk scale height $H$ is set to $0.05$ giving a temperature $T_0$ at
  $1$ AU from the star of about $620$ K. 
  Since during the stellar encounter the disks may become warped, the
  isothermal temperature profile is always computed respect to the average
  median plane of the disk.
  Each disk is simulated with $850000$ SPH particles for a total of $1.7$
  million particles for the whole system.
   
  We adopted the standard SPH artificial viscosity \citep{monaghan83}, 
  introducing a linear ($\alpha_{\rm SPH}$) and a quadratic 
  ($\beta_{\rm SPH}$) term whose initial values are set to $0.1$ and $0.2$, 
  respectively.
  \cite{meru12} showed that these terms can be compared to the 
  \cite{shakura73} $\alpha$--viscosity parameter $\alpha_{\rm SS}$ 
  defined as 
  $\nu = \alpha_{\rm SS} c_{\rm s} H$ where $\nu$ is the kinematic viscosity,
  $c_{\rm s}$ is the sound speed and H is the vertical
  pressure scale height.
  Using their relations we find that, away from the close approach, the
  corresponding $\alpha_{\rm SS}$ in our models is about $0.002$.

  One disk harbors a Jupiter size planet orbiting at $18$ AU on a circular
  orbit not inclined respect to the disk. 
  We set the planet on an outside trajectory since we want to maximize the
  effects of the stellar encounter in order to test if, in spite of 
  the highly
  perturbing configuration, the disk is still able to damp the planet 
  eccentricity and inclination after the flyby. 
  The second star is started on a hyperbolic orbit with a minimum impact
  parameter $q$ fixed at the beginning of the simulation.
  The initial distance between the two stars is $450$ AU (approximately
  $500$ yr before the encounter) and the relative velocity at infinity is
  set to $1$ km/s, a typical value in clusters.
  Different inclinations are considered for the passing star to test the
  amount of disk warping and planet excitation in inclination.
  The two disks are oriented at different angles up to the extreme case of
  prograde-retrograde encountering disks where one is rotating in the 
  direction of the encounter orbit, and the other against it. 

  Prior to the encounter, the planet begins to carve a gap. 
  We might have let the disk--planet system to evolve longer in order to
  allow the planet to create a full gap.
  However, as also shown in \cite{frane09} and \cite{marpi13}, a close
  stellar encounter strongly perturbs the disk and planetary orbit, 
  cancelling out any pre-existing structure including a gap around the 
  planet.
  A new gap is created some time after the stellar flyby when the disk 
  relaxes, as shown in our simulations.  

\section{Definition of the different flyby geometry and disk inclinations}
\label{setup}

  Due to the large amount of CPU time required by each 3D run, we have
  performed a limited number of simulations covering, by necessity, a 
  small but meaningful portion of the initial phase space.
  We look for the most perturbed configurations since our goal is to test
  the ability of the gaseous disk to damp the planet orbital parameters
  after the stellar encounter even in extreme conditions. 
  In Tab.~\ref{numb} we summarize the main initial 
  parameters of each simulation (first 5 columns of the table) while 
  the changes of some relevant disk properties, during the 
  evolution of the system, are given in the other columns.
  Their meaning will be described later on.

  To compute the initial set up for the disks and the hyperbolic 
  encounter, we start from a coplanar configuration where the 
  hyperbolic trajectory of the second star has the pericenter on the 
  positive $x$--axis while the star harbouring the planet is at the origin
  of the $x$--$y$ frame. 
  We then tilt the hyperbolic orbit around the $x$--axis by an angle 
  $i_{\rm s2}$ to vary the approach geometry.
  We also rotate the disk of the second star around the $y$--axis by an
  angle $i_{\rm d2}$.
  The disk of the primary star has instead its median plane on the 
  $x$--$y$ plane. 
  We could have further varied the initial configuration by rotating both
  the disks around the $x$ and $z$--axis, but due to the heavy CPU load of
  each simulation, we considered only the rotations defined by 
  $i_{\rm s2}$ and $i_{\rm d2}$. 

\begin{figure}[htbp!]
    \centering
    \includegraphics[width=0.55\textwidth]{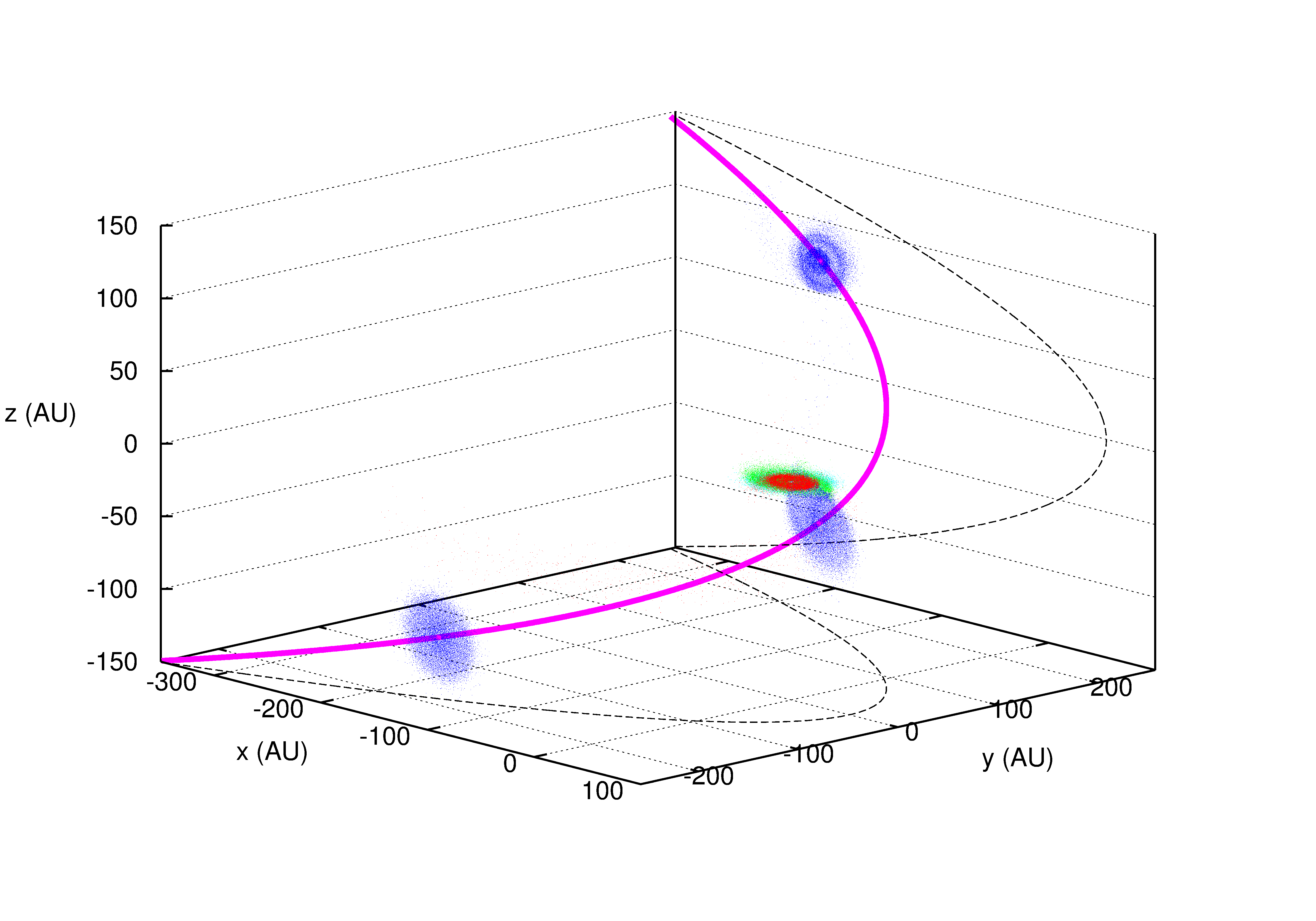}
    \includegraphics[width=0.5\textwidth]{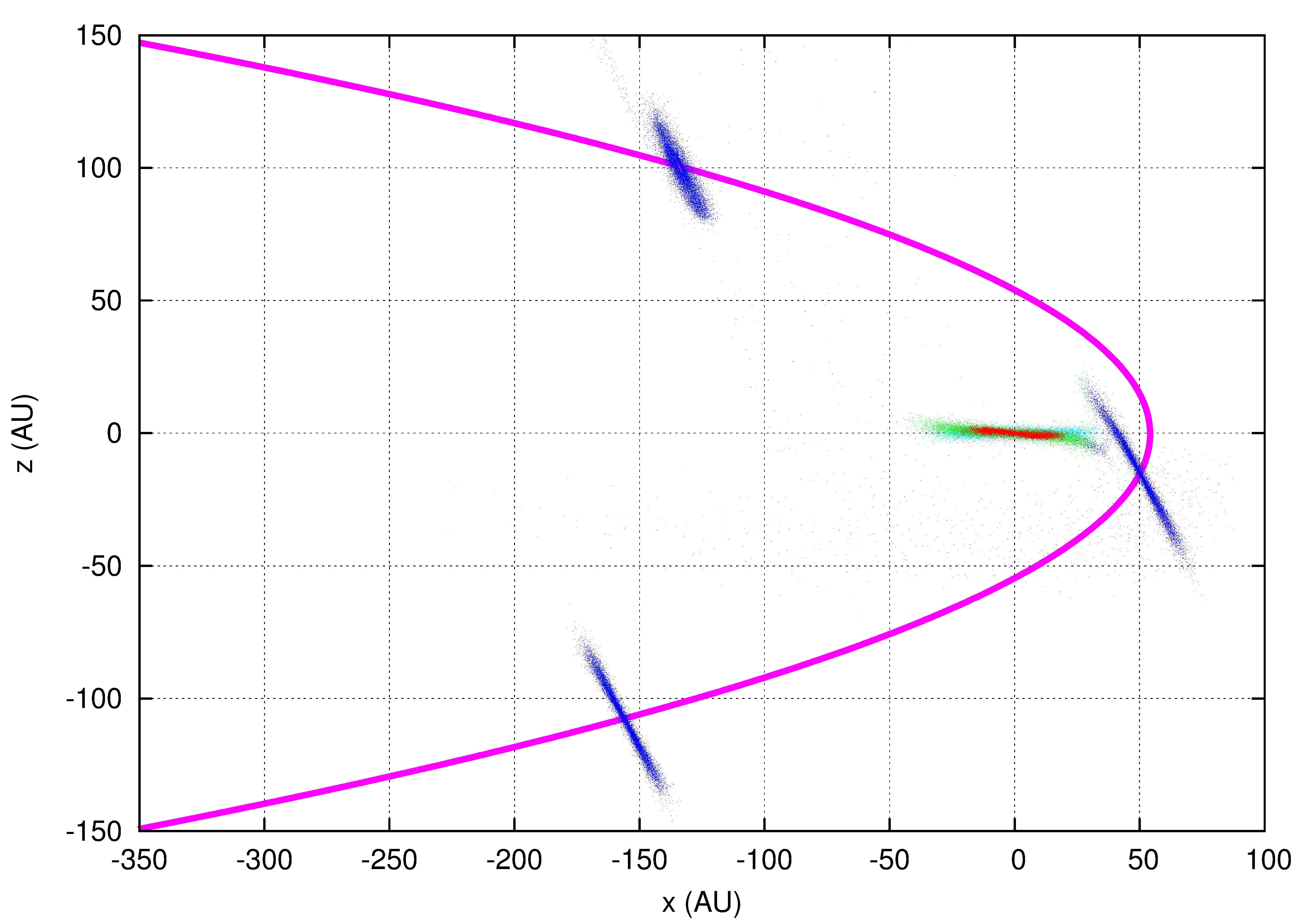}
    \caption{ In the top panel we illustrate in 3D the encounter 
    geometry of model h. 
    The hyperbolic trajectory is plotted as a continuous line while its 
    projections on the (x,y) and (x,z) plane are shown as dashed lines.
    The disks around the primary and secondary star are shown at 3 
    different evolutionary times around the pericenter passage.
    The primary disk is at the center of the reference frame so its shape
    overlaps at the 3 different times.
    In the bottom panel we show a projection of the 3D plot on the (x,z)
    plane, a view that helps to understand the dynamical configuration of
    the encounter.}
    \label{fig0}
  \end{figure}

  In Fig.~\ref{fig0}a we show an example of this procedure giving a 
  full 3D
  view of the encounter geometry for the case h where the disk is inclined
  of $60^o$ respect to the primary disk and the hyperbolic trajectory is
  inclined of $30^o$ respect to the primary disk plane.
  The hyperbolic trajectory is shown during the
  pericenter passage and the two disks are plotted at 3 different
  evolutionary times.
  A 2D projection in the (x,z) plane (Fig.~\ref{fig0}b) of the 
  previous plot gives an additional
  view of the mutual orientations of stars and disks.

  In a test case we also changed the minimum approach distance $q$ (case
  b) and we also performed a simulation with no disk around the
  second star to compare with all other cases with two disks (case d).
  This model is relevant for understanding the role of the disk around the
  second star in the evolution of the planet. 

\section{Cases a, b and c: $i_{\rm d2} = 45^o$.}
\begin{figure}[htbp!]
    \centering
    \includegraphics[width=0.46\textwidth]{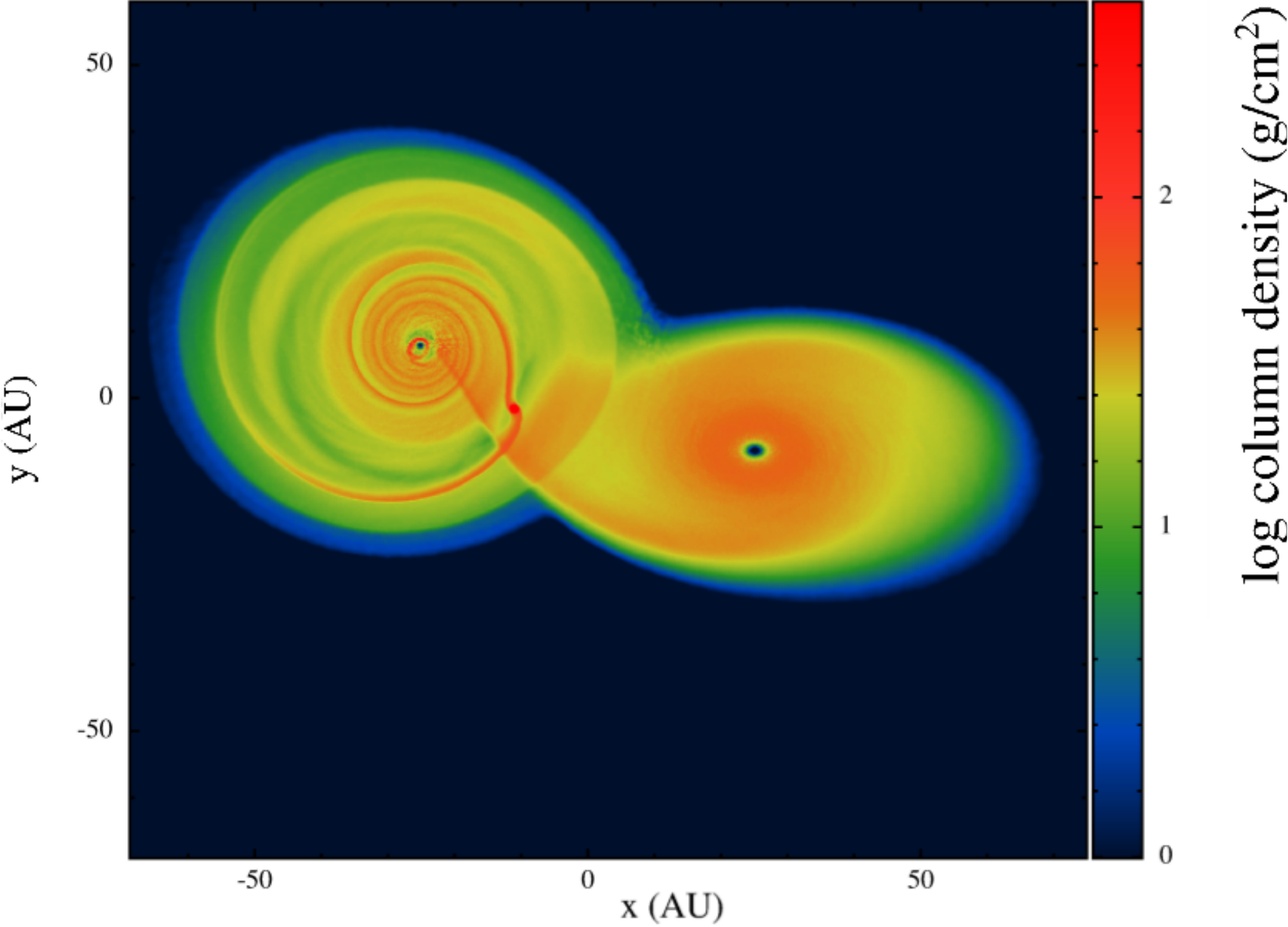}
    \includegraphics[width=0.46\textwidth]{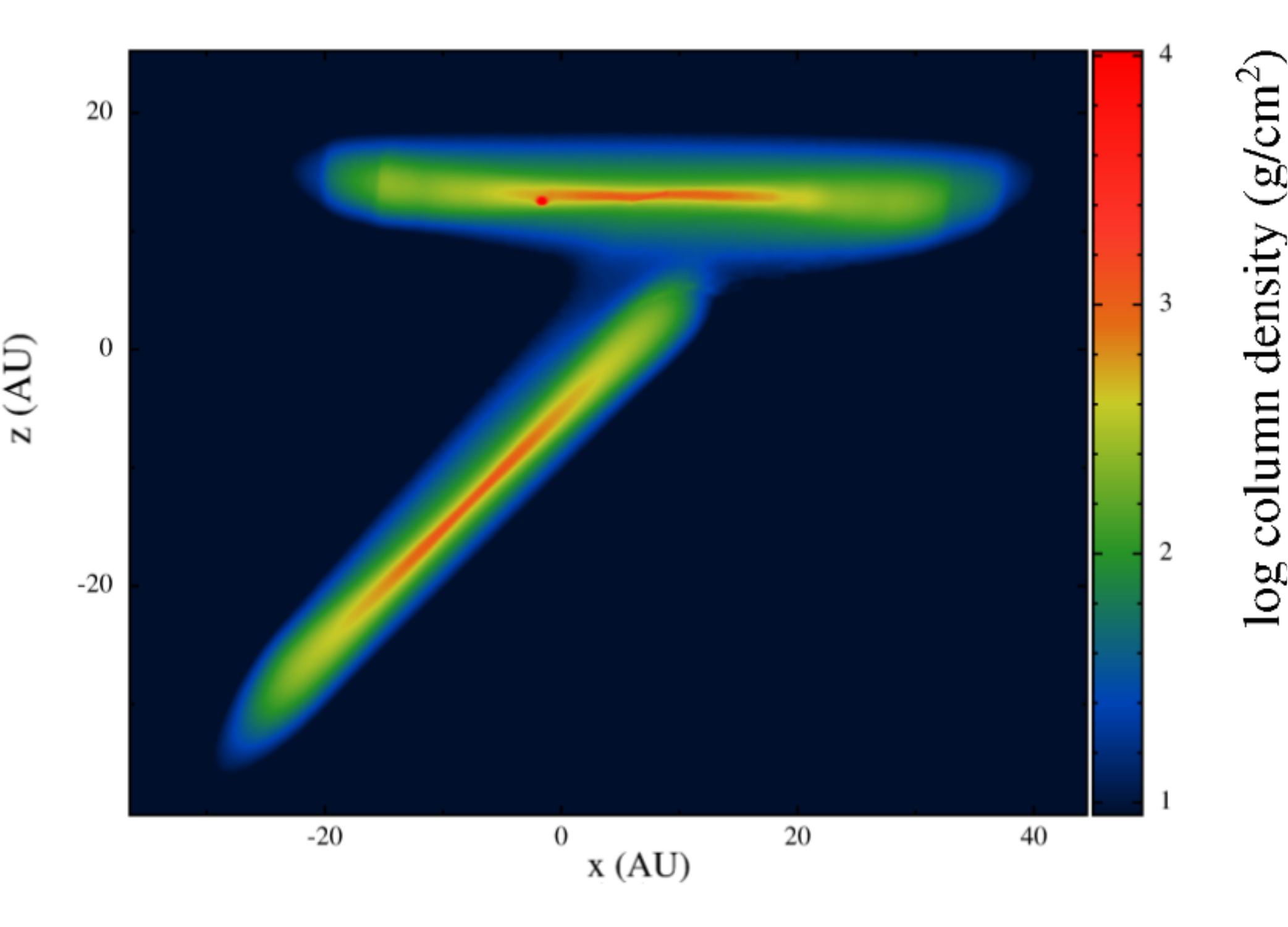}
    \includegraphics[width=0.46\textwidth]{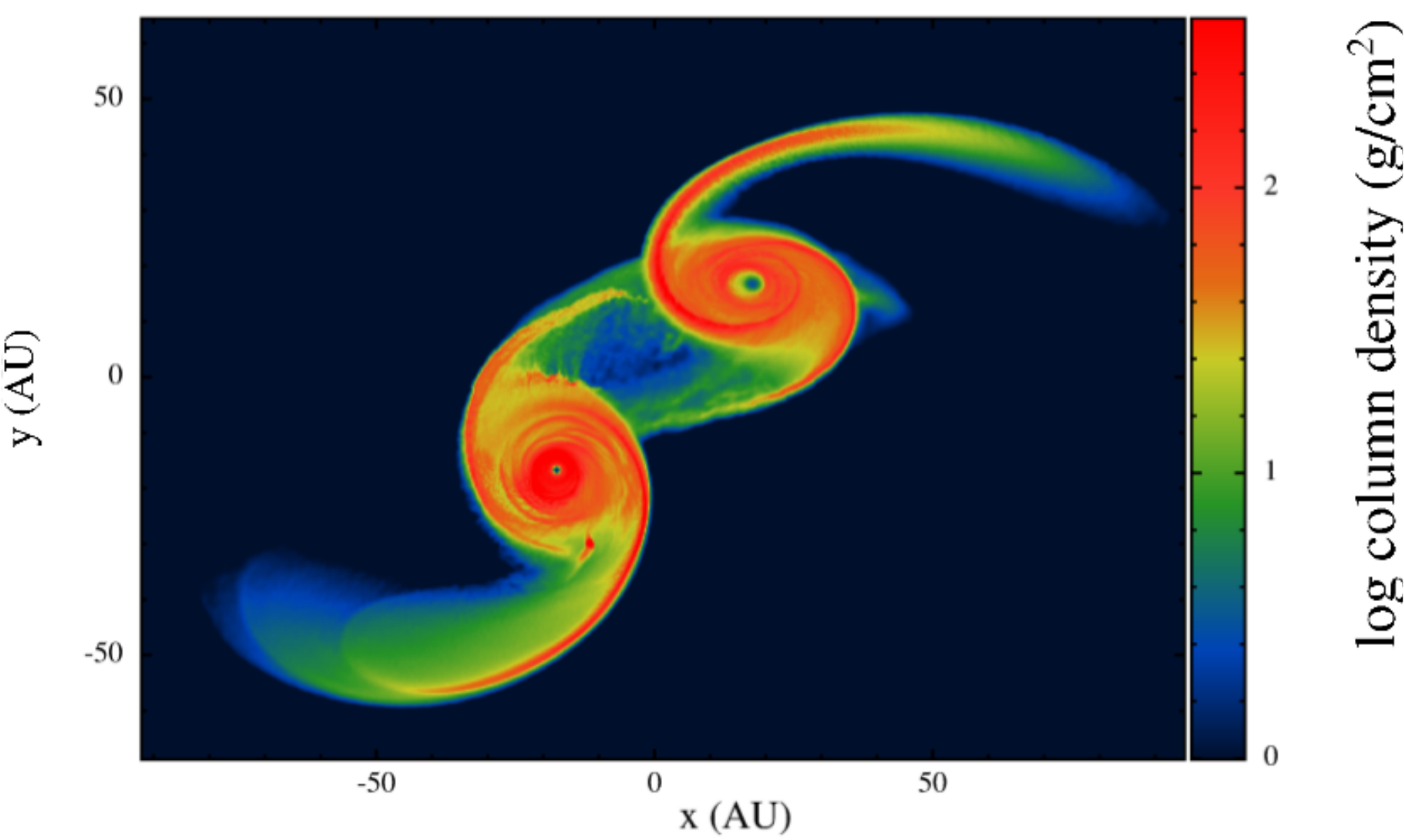}
    \includegraphics[width=0.46\textwidth]{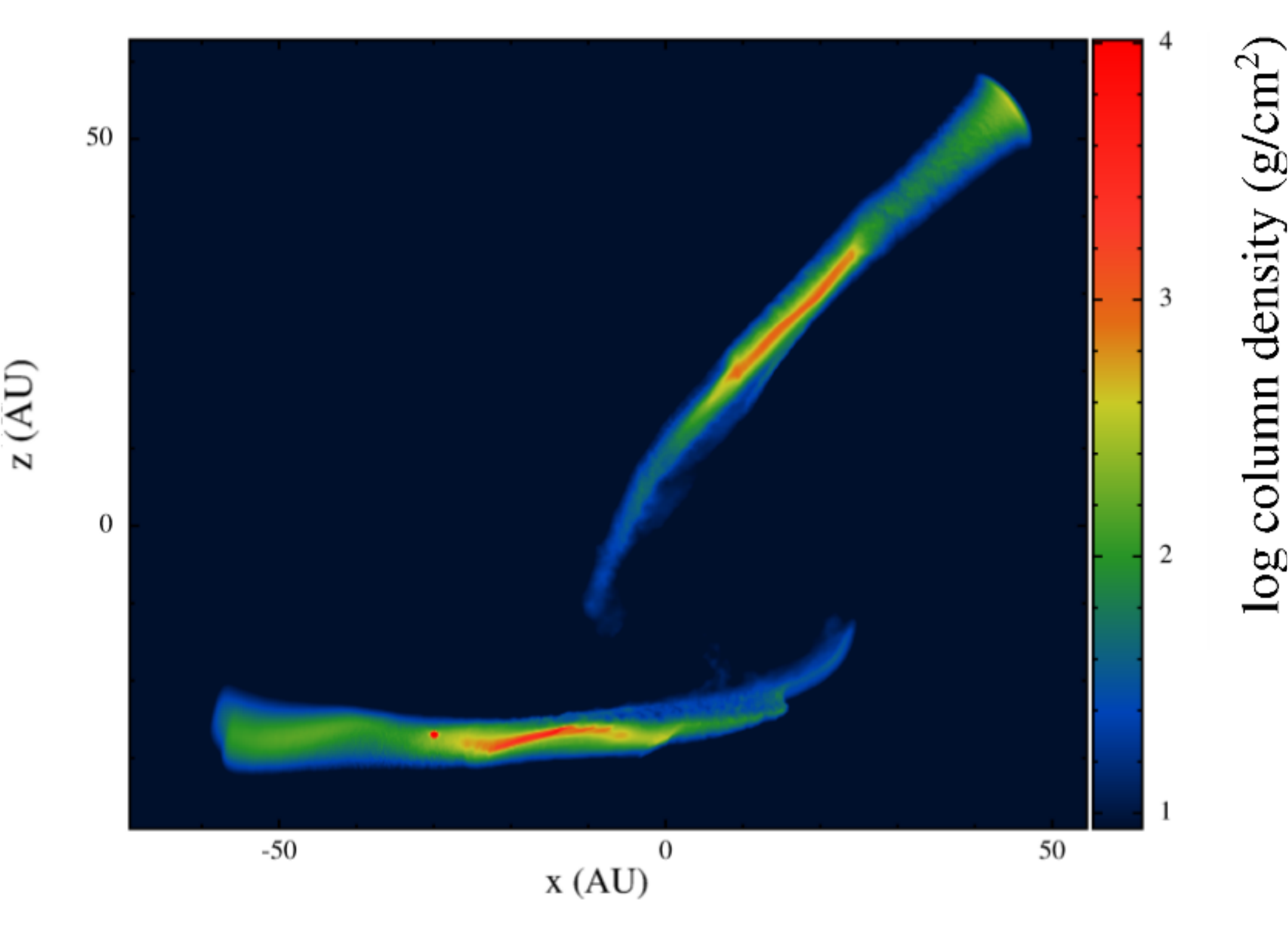}
    \caption{Logarithm of the superficial density (column integrated) of
      the circumstellar disks in the $x$--$y$ and $x$--$z$ plane, at $2$
      different evolutionary steps during the stellar flyby for the 
      model c. Note that the color scale for the 
      superficial density $\Sigma$ is different in the 
      $x$--$y$ (face--on)  and $x$--$z$ (edge--on)  
      projections since the bulk density 
      $\rho$ is integrated 
      along different paths.
      The top plots show the disks when the stars are close to the
      pericenter passage, the bottom plots when the stars are departing
      just $60$ yrs after the encounter.}
    \label{fig1}
  \end{figure}

\begin{figure}[htbp!]
    \centering
    \includegraphics[width=0.5\textwidth]{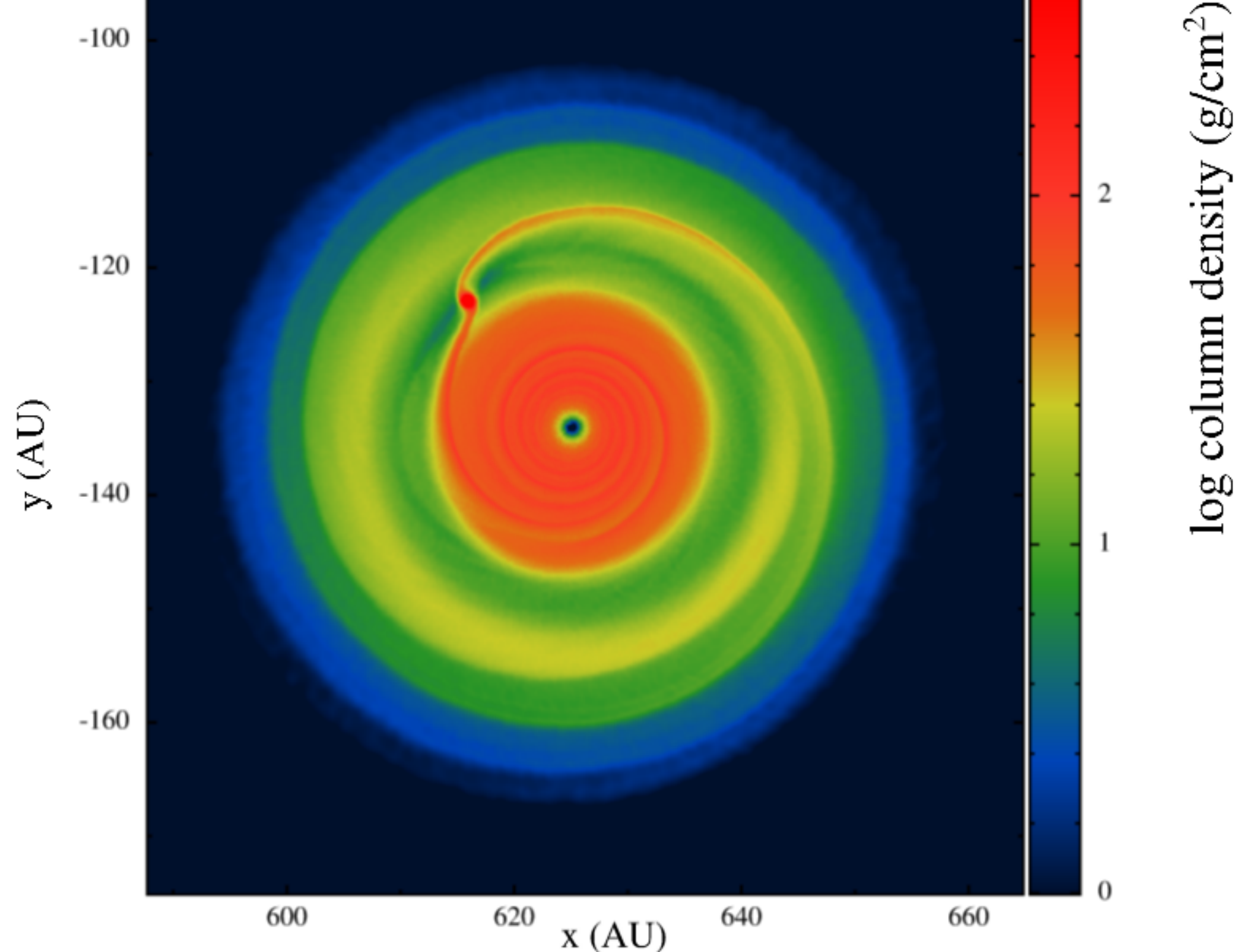}
    \includegraphics[width=0.5\textwidth]{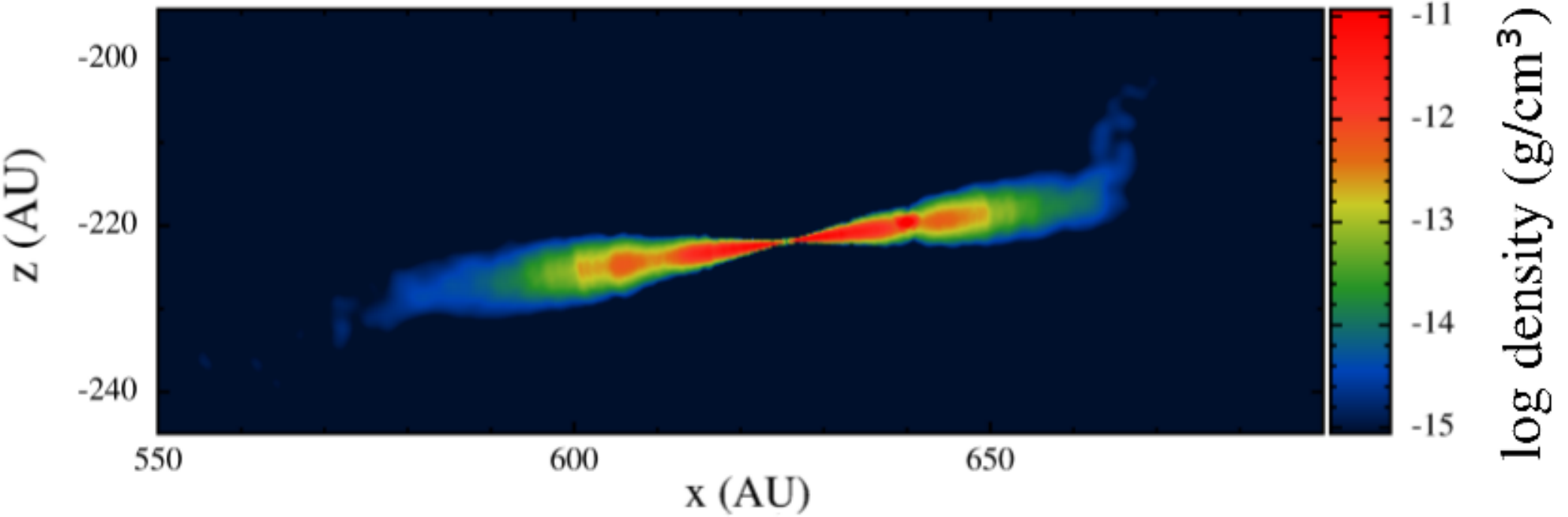}
    \caption{Logarithm of the superficial density (column integrated) of
      the disk around the star with planet and its $x$--$z$ slice $2800$
      yrs after the flyby for the model c. 
      The disk has reached a quiet state and the planet has almost
      completely carved a gap around its orbit.}
    \label{fig1x}
  \end{figure}

  We first analyze the models where the disks are mutually inclined of
  $45^o$ i.e. $i_{\rm d2}=45^o$.
  In two cases with this configuration 
  we test the dependence on the periapse distance $q$ of the
  hyperbolic orbit on the evolution of the disks and planet orbit (cases 
  a and b).
  In the third case c we incline also the hyperbolic trajectory setting
  $i_{\rm s2} = 60 ^o$.
  This last case is more general in terms of geometric configuration
  and we will adopt it as reference case comparing its outcome to the
  other two. 
  In Fig.~\ref{fig1} the evolution of the disks and planet are shown at 
  different times for case c.
  At the minimum distance (first two plots) the disks begin to interact
  becoming highly eccentric and strongly warped.
  The outer edge of the disk around the passing star overlaps close to the
  center with that of the primary star. 
  At subsequent times, the disks are very distorted, and tidal tails are
  launched causing a significant mass exchange. 
  It is noteworthy that in the third plot the disk around the passing
  star (secondary star) appears very eccentric, but this is a perspective
  effect.
  When the integrated density is computed, the inclined disk is projected
  on the $x$--$y$ plane and so it appears more eccentric due to its
  inclination. 
  About $2800$ yr after the flyby, the primary disk is returned to a quiet
  state and the planet has carved a new gap in the disk, as illustrated in
  Fig.~\ref{fig1x}.

  \subsection{Evolution of the planet orbit}

    The planet in case c, which appears in the plots of Fig.~\ref{fig1} 
    and Fig.~\ref{fig1x} as a local
    overdensity, is moved on an inclined and eccentric orbit after the 
    stellar flyby due to the gravitational pull of the passing star. 
    However, the subsequent interaction with the disk damps both the 
    eccentricity and inclination and, on a short timescale,
    the planet is driven back on an almost circular orbit not inclined
    respect to the disk plane.
    All the effects of the stellar flyby have been erased apart from a
    shift in semimajor axis. 
    However, the final planet orbit may be slightly inclined 
    respect to the equatorial plane of
    the primary star since the disk, after a warped phase, relaxes into an
    inclined configuration (see Fig.~\ref{fig1x}, bottom plot). The planet,
    tidally interacting with the disk, 
    is dragged back not to the star equatorial plane but to
    the disk median plane, slightly tilted, after the flyby, 
    respect to the star 
    equatorial plane. 

\begin{figure}[htbp!]
      \centering
      \includegraphics[width=0.5\textwidth]{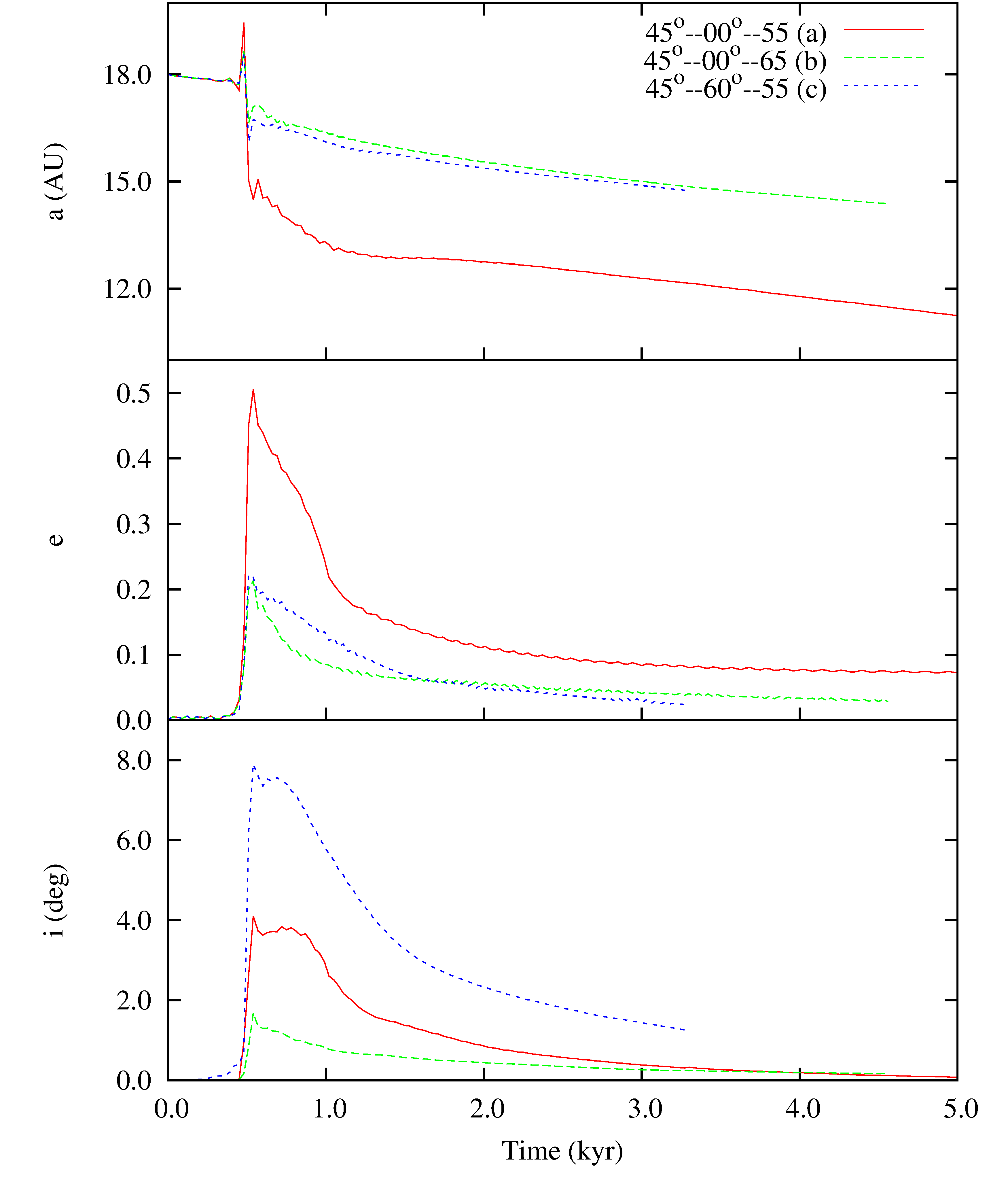}
      \caption{Evolution of the semimajor axis (top panel), 
        eccentricity (middle panel), and inclination (bottom panel) of the
        planet in models a,b, and c.} 
      \label{fig3}
    \end{figure}

    In Fig.~\ref{fig3} we show the time evolution of the orbital 
    parameters of the planet in the $3$ simulations a, b, and c.
    The semimajor axis, after a jump caused by the stellar encounter,
    decreases with time due to the tidal interaction with the disk.
    A new gap around its orbit is forming, as shown in Fig.~\ref{fig1x}
    (top plot) for the case c ($i_{\rm d2} = 45^o$ and 
    $i_{\rm s2} = 60^o$), but it is not yet deep enough to cause the slow
    type I migration which will probably start at later times. 
    The delay is also due to the initially inclined orbit of the planet
    respect to the disk which slows down the gap carving. 

    The eccentricity pumped up by the stellar flyby is quickly damped when
    the second star departs from the primary after the encounter.
    This is illustrated in Fig.~\ref{fig3} middle panel. 
    In the most perturbed configuration (case a) the eccentricity grows
    up to $0.5$ and is quickly reduced to $0.2$ in less than $1000$ yr.
    After this initial phase, the damping is slower occurring at a rate of
    about $6\times10^{-6}$ yr$^{-1}$ leading to a full damping on a
    timescale, in the worst case, of a few $10^4$ yr.
    In cases b and c the initial eccentricity excitation is of the
    same order of magnitude.
    The inclination of the hyperbolic orbit of case c possibly reduces
    the strength of the passing star perturbations on the planet
    eccentricity as it does the larger flyby minimum distance in case b.
    However, the damping appears to be faster in case c where the final
    eccentricity, after $2700$ yr from the flyby, is only $0.02$.
     
\begin{figure}[htbp!]
      \centering
      \includegraphics[width=0.5\textwidth]{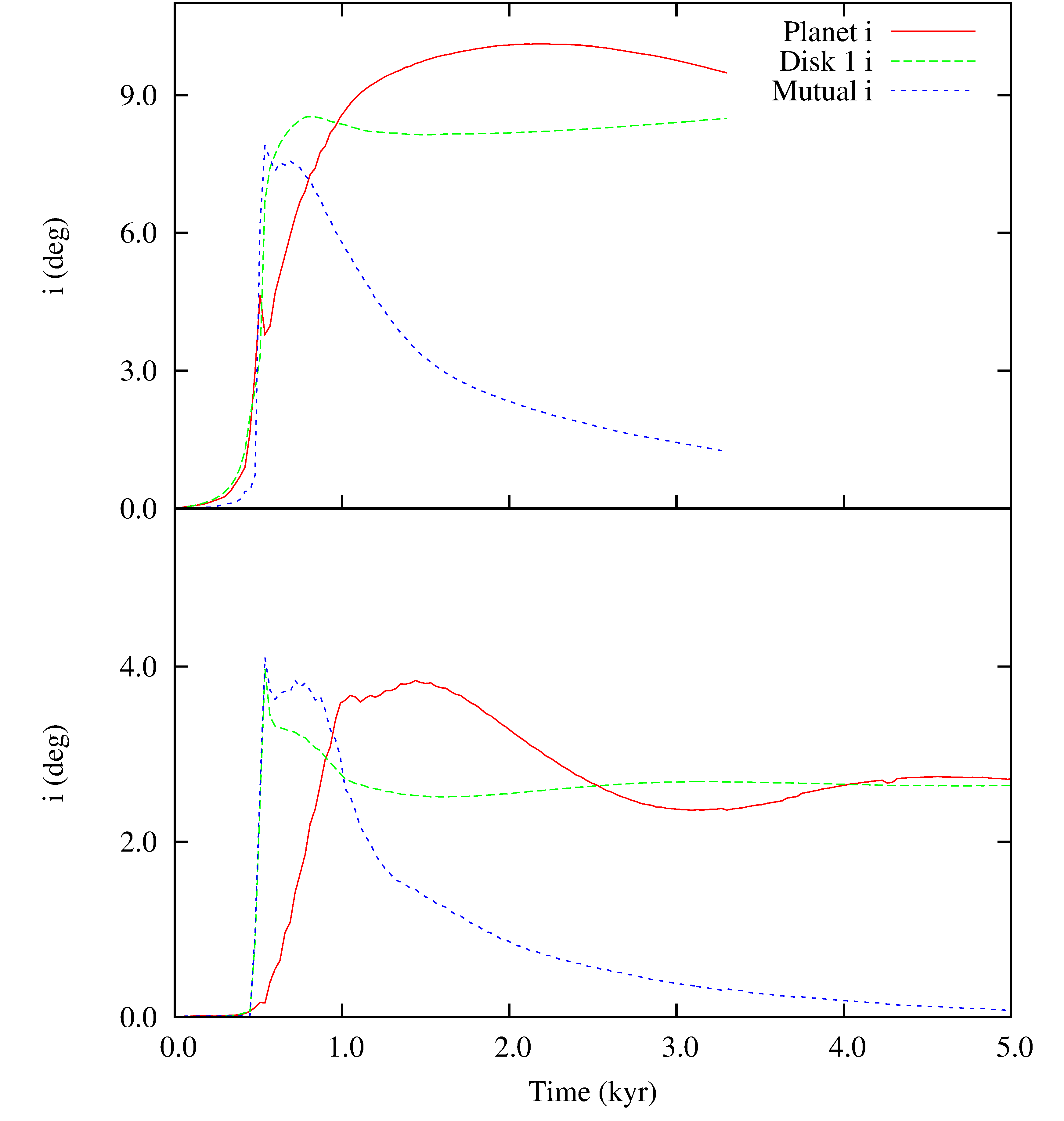}
      \caption{Evolution of the disk, planet and mutual inclination for
        cases c, top panel, and a, bottom panel.}
      \label{fig3b}
    \end{figure}

    In case c the inclination of the planet respect to the equatorial
    plane of its host star is excited during the flyby up to about
    $4.5^o$.
    After the stellar encounter also the disk is inclined respect to its
    initial plane and its inclination is about $8.5^o$ as shown in
    Fig.~\ref{fig3b}, top plot. 
    The gravitational interaction between the planet and disk drives the
    planet back into the disk and its inclination is slowly brought to the
    same value of that of the disk.
    The mutual inclination between the planet and the disk decreases  
    smoothly with time to a low value of about $1^o$ when the simulation
    was stopped after $2.3$ kyr.
    In Fig.~\ref{fig3b}, bottom plot, we show the evolution of the disk
    and planet inclinations in the case a.
    The orbit of the planet is not excited in inclination by the stellar
    flyby since $i_{\rm s2} = 0$.
    However, the disk around the primary star is first warped and then it
    settles down to an inclination of about $2.8^o$.
    The planet inclination grows until the mutual inclination between
    planet and disk is negligible.  

    The outcome of these models confirms that the presence of the disk 
    leads to a fast damping of the eccentricity and inclination due to the
    stellar flyby, erasing any trace of the event in the trajectory of the
    planet.
    Such an event might be recorded in the inclination of the disk (and
    then of the planet) respect to the equatorial plane of the star 
    but the expected tilting would be very small.
    It could accelerate the inward migration of the planet by
    restarting type I migration after the stellar flyby when the gap is
    erased but soon a new gap would be cleared and type II migration 
    restored.  

\section{Coplanar cases d, e, and f}

\begin{figure}[htbp!]
    \centering
    \includegraphics[width=0.5\textwidth]{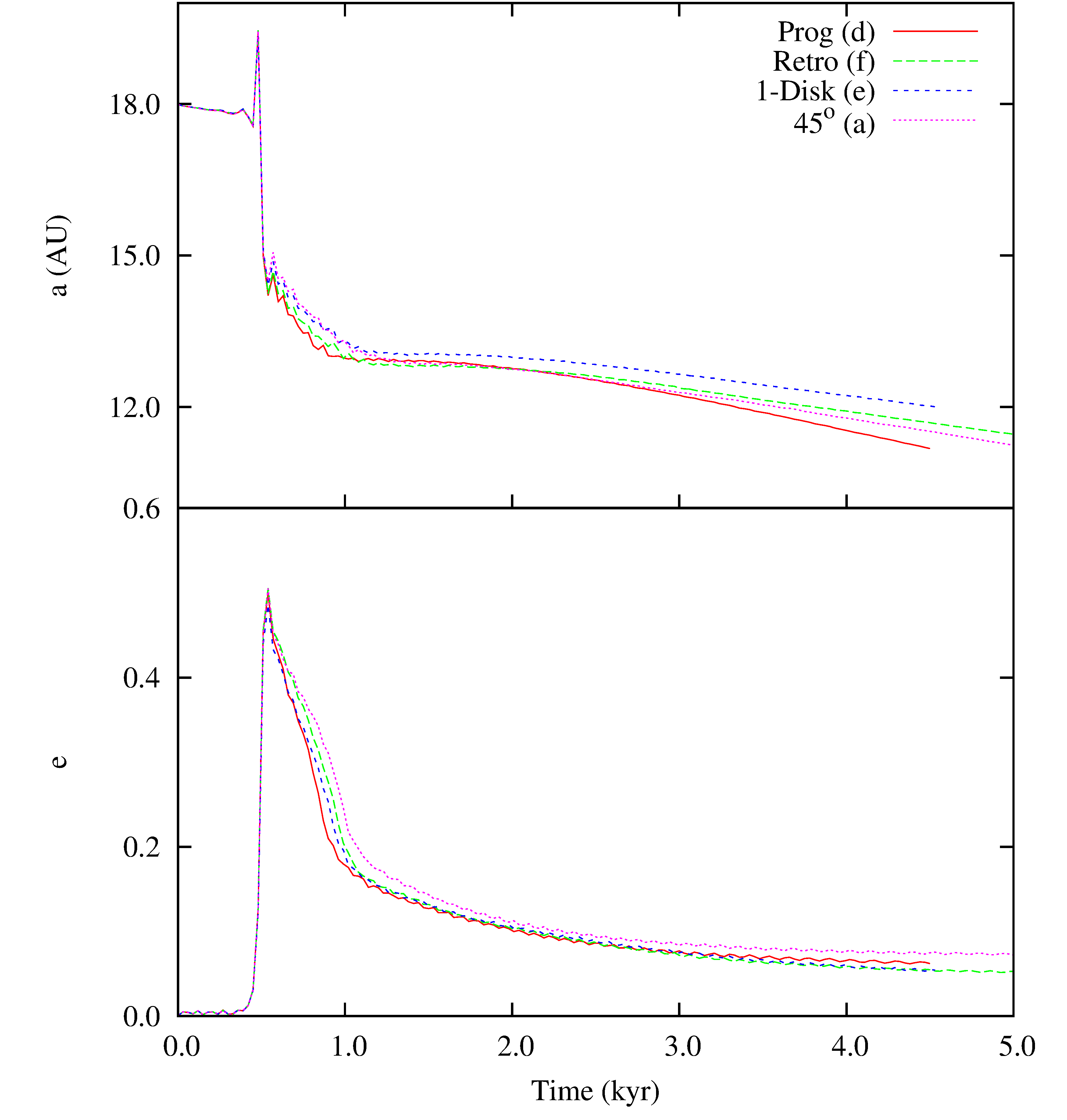}
    \caption{Evolution of planet's orbital elements $a$ and $e$ in models
      d, e, and f (coplanar cases), compared with the case a.} 
    \label{fig4x}
  \end{figure}

  These models assume a coplanar initial configuration for the disks and for
  the hyperbolic orbit of the passing star, i.e. $i_{\rm d2} = 0^o$ and
  $i_{\rm s2} = 0^o$.
  They are the analogous of 2D models except that the disks are computed
  in 3D. 
  We do not have changes in inclination and we concentrate only on the
  semimajor axis and eccentricity variations.
  As illustrated in Fig.~\ref{fig4x}, the eccentricity and semimajor axis
  jumps in the three cases d, e, and f are about the same.
  This confirms that the gravitational perturbations of the passing star
  dominates the dynamical evolution of the planet during the flyby.
  The subsequent damping is not significantly influenced by the different
  evolution and mass loss of the disk (see Tab.~\ref{numb}) but it is
  slightly faster compared to that of case a ($i_{\rm d2} = 45^o$,
  $i_{\rm s2} = 0^o$), shown as a reference in Fig.~\ref{fig4x}.
  This is possibly due to the disk--planet mutual inclination that, in
  case a, reduces the tidal interaction. 

  We considered the prograde-retrograde encountering configuration as an
  extreme case with one disk (that hosting the planet) rotating in the
  direction of the encounter orbit, and the other rotating against the
  encounter.
  The star is on a hyperbolic orbit coplanar to both disks.
  The evolution of the disks is not symmetric since for one disk (the
  primary) the secondary star is on a prograde trajectory while for the
  second disk the primary star is seen as moving on a retrograde
  trajectory. 
  In Fig.~\ref{fig4} the disk integrated column density is shown shortly
  after the stellar flyby (top panel) and $60$ years later (bottom 
  panel). 
  The two disks are not perturbed in a symmetric way, as for the 
  coplanar prograde case, and only the primary shows a consistent 
  mass loss along the trailing tidal wave. 
  The disk around the secondary passing star does not become significantly
  elliptic and it does not develop tidal waves before the encounter.
  It appears to be less perturbed by the primary star gravity and this is
  in agreement with the findings of \cite{hall96} claiming that the 
  angular momentum transfer between a perturber and a disk in a retrograde
  configuration is smaller.
  A similar effect was also observed in \cite{forgan09} for more massive
  disks.
  At the pericenter of the flyby the two disks come in contact and mass is
  transferred from one to the other. 
  As it can be noted in Fig.~\ref{fig4}, the primary disk shows the usual
  spiral structure with the trailing arm extending into the secondary
  disk. 
  Due to the direct interaction between the disks, the upper part of the
  secondary disk (retrograde respect to the primary star) is strongly
  distorted while the bottom part of the disk is almost circular and 
  unperturbed.
  According to Tab.~\ref{numb} the primary disk loses approximately $1/4$
  of its mass while the secondary disk gains $2\%$ of its initial mass.  

  In spite of the strong differences in the disk evolution between the
  prograde and retrograde cases, the orbital evolution of the planet is
  not significantly different.
  This reinforces the idea that the tidal damping of the planet 
  eccentricity is a robust mechanism that does not depend significantly on
  the disk evolution.
  Once the perturbing effects of the star flyby are passed, and this 
  occurs on a short timescale, the disk relaxes and drives the planet back
  to a circular non--inclined orbit. 

  \subsection{Mass exchanged by disks or lost to infinity}

    A significant difference between the 2D models of \cite{marpi13} and
    the 3D results presented here is the amount of mass lost by the disk
    with planet during the encounter.
    In \cite{marpi13} the primary disk retained about half of its initial
    mass at the end of the simulation while in the 3D cases here presented 
    the fraction of mass loss, summarized for all models in 
    Tab.~\ref{numb}, is significantly smaller and it strongly depends on
    the encounter geometry, as shown in Fig.\ref{fig1M}. 
    This difference may be ascribed to the following reasons.
    \begin{itemize}
      \item In previous 2D models only one disk was included while in the 
        present models we consider 2 disks. The presence of the 
        disk around the secondary star has important consequences 
        for the evolution of the system
        since a consistent fraction of the mass lost during the 
        encounter by the disk around the primary is 
        replaced by mass stripped from the disk of the secondary.
        According to Tab.~\ref{numb} up to $13\%$ of the disk initial
        mass can be exchanged between the two disks.
        Explicative in this context is the outcome of model e.
        In Fig.\ref{fig2} we show the 
        density distribution after the stellar encounter illustrating 
        how, during the stellar flyby, the second star, initially without its
        own disk, strips material from the primary. The new captured 
        disk has a mass 
        with mass equal to $15\%$ of the initial disk mass of the primary
        (see Tab.~\ref{numb}).
        This proves that a large amount of mass can be transferred 
        from one disk to the other, an effect not observed 
        in \cite{marpi13} because the
        secondary star was out of the computational grid. 
      \item In 3D the disk evolution, and then the mass loss, depends also
        on the behaviour along the vertical direction. This leads to 
        a lower mass loss in 3D models respect to 2D ones as illustrated
        in Fig.\ref{fig1bis} where we compare the outcome of the 3D model 
        e with a 2D simulation,
        performed with the SPH code, 
        having the same initial physical parameters and profiles of model e. 
        The mass loss is 
        slightly higher in the 2D model just after the stellar flyby. 
        However, the comparison also suggests 
        that changing the dimensions of the simulations has 
        less impact on the mass loss than the existence of a
        secondary disk. As a consequence, the mass loss difference 
        between 2D and 3D models
        is mainly due to the mass exchange between the 
        two circumstellar disks.
        Note that the  
        apparent increase of the disk mass after it hit the minimum is 
        fictitious and it is due to the difficulty of defining the disk 
        borders during the encounter
        when the shape of the disk is very elongated and asymmetrical.
        \begin{figure}[htbp!]
          \centering
          \includegraphics[width=0.5\textwidth]{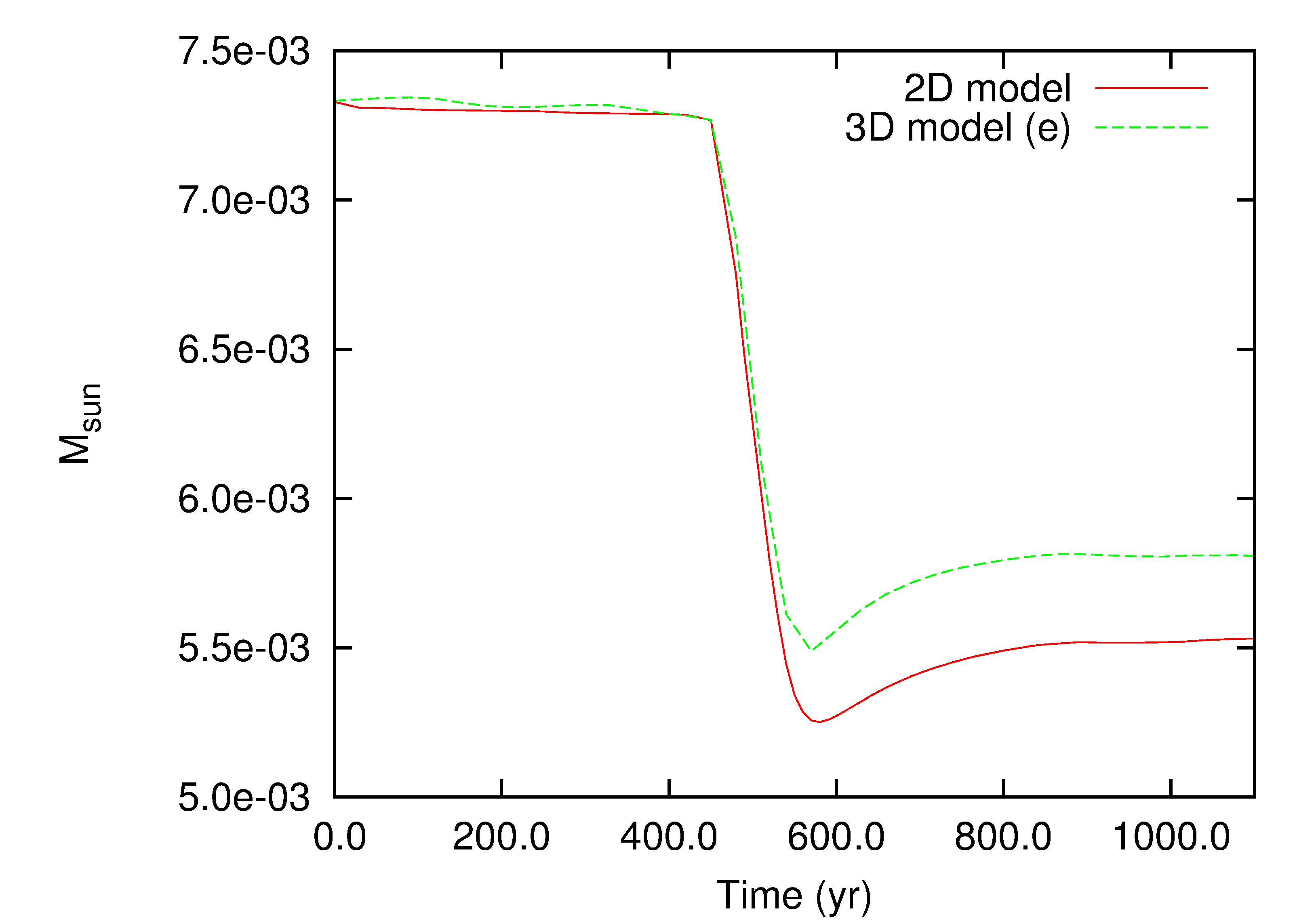}
          \caption{ Disk mass evolution in the 3D model e compared
            to a 2D model with identical initial parameters and integrated
            with the same SPH code VINE. The mass loss
            during the encounter is higher in the 2D model.}
          \label{fig1bis}
        \end{figure}
      \item Previously published 2D simulations were performed with a grid
        code having a limited spatial domain centered on the primary star. 
        Mass exiting the grid domain was lost by the system.
        In the models shown here an SPH code is used and there are no 
        boundaries for the disks.
        As a consequence, gas excited on highly eccentric trajectories
        during the stellar flyby is allowed to fall back on the disk at
        later times while in the simulations with the grid code it was
        lost.
        This can be noted also in Fig.~\ref{fig1M} where the mass of the
        disk is computed adding up all the mass moving within 
        $R_{\rm out} = 50$ AU from the star.
        During the flyby mass is lost but part of it comes back within
        $R_{\rm out}$ before the onset of an almost constant mass loss due
        to viscosity, spiral waves excited by the planet and residual disk
        instability related to the stellar flyby. 
      \item The 2D models were propagated a bit longer in time and this
        might explain part of the larger mass loss. 
        In the 3D models the disk is 
        still loosing a small amount of mass when we halt the simulation.
        This last effect, however, accounts for only a small fraction of 
        the difference in the mass loss between 2D and 3D models. 
    \end{itemize}

    \begin{figure}[htbp!]
      \centering
      \includegraphics[width=0.5\textwidth]{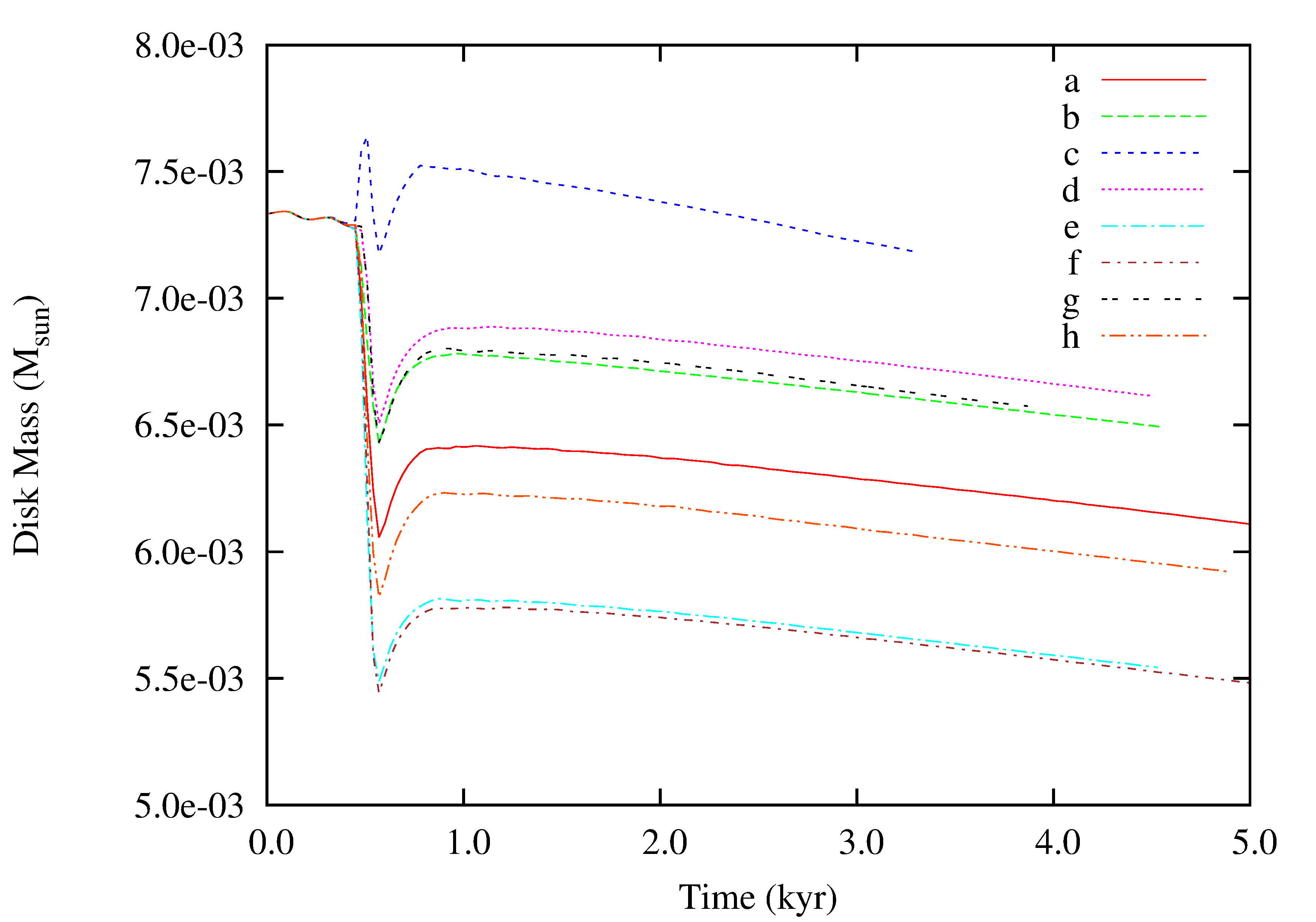}
      \caption{Mass of the disk around the star harbouring the planet as a
        function of time for the different models summarized in 
        Tab.~\ref{numb}.
        The flyby is marked by a jump followed by a slow mass decrease.}
      \label{fig1M}
    \end{figure}

   Since the aim of our calculations is to show that the disk damps 
   the excitation
   caused by the stellar flyby on the planet orbital elements, we are 
   interested in 
   the mass loss only occurring during the damping.  
   The disk, in this short timescale ($\sim$ 5 kyr), 
   possibly does not reach a configuration where
   the mass loss is only due to viscosity and spiral waves excited by the
   planet but it is still 
   loosing mass because of the binary perturbations. 
   However, this is not relevant for our
   computations since the damping of the planet orbit has already occurred.

    The mass exchange might cause a limited variation in the original 
    metallicity of the disk compared to that of the host star assuming 
    that the two encountering stars have significantly different values of
    metallicity.
    However, the amount of pollution is tiny and possibly without 
    significant consequences for planet formation.
    The reduced mass loss in the more realistic 3D models presented here
    leads to a longer lifetime of the disk and a stronger damping effects
    on the eccentricity and inclination of the planet after the flyby.  

    \begin{figure}[htbp!]
      \centering
      \includegraphics[width=0.5\textwidth]{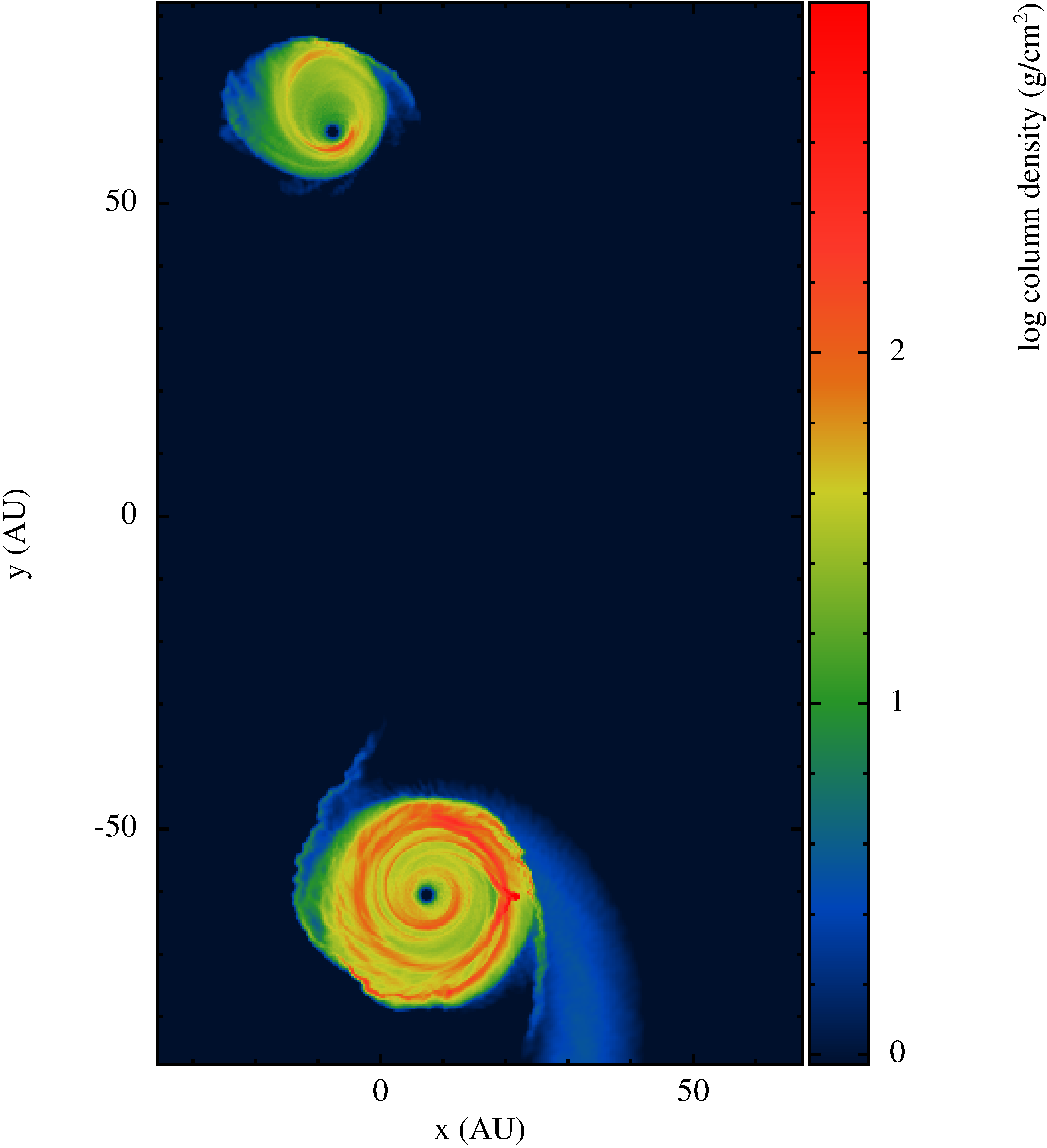}
      \caption{Capture of a disk by the second star, initially without any
        disk  (model e), during the stellar flyby.
        The plot show the system shortly after the stellar encounter when the 
        material has been transferred from one star to the other via the 
        extended tidal wave.} 
      \label{fig2}
    \end{figure}

  \begin{table*}
    \begin{center}
    \caption{List of models} \label{numb} 
    \begin{tabular}{c c c c c c c c c c c c c}
      \toprule[1.5pt]
      \bf id & $i_{\rm d2}$ & $i_{\rm s2}$ & $q$ (AU) & 
        nd & $\frac{\Delta M_{\rm d1}}{M_{\rm d1}}$ & 
        $\frac{\Delta M_{\rm d2}}{M_{\rm d2}}$ & $M_{12}$ & 
        $M_{21}$ & $e_{\rm d1}$ & $e_{\rm d2}$ & 
        $i_{\rm d1}$ & $i_{\rm d2}$ \\
        \midrule
      a  & $45^o$ & $0^o$ & $55$ & $2$ & $-16\%$ & $-3\%$ & $11\%$ & 
        $10\%$ & $0.09$ & $0.09$ & $2.6^o$ & $-3.4^o$ \\
        \midrule
      b  & $45^o$ & $0^o$ & $65$ & $2$ & $-11\%$ & $-1\%$  & $7\%$  & 
        $6\%$ & $0.05$  & $0.05$ & $1.0^o$ & $-1.8^o$ \\
        \midrule
      c  & $45^o$ & $60^o$ & $55$ & $2$ & $-2\%$  & $-17\%$ & $3\%$  & 
        $13\%$  & $0.04$  & $0.07$ & $8.5^o$ & $2.0^o$ \\
        \midrule
      d  & $0^o$ & $0^o$ & $55$ & $2$ & $-6\%$  & $-5\%$ & $8\%$  & $11\%$  
        & $0.08$  & $0.09$ & $0.0^o$ & $0.0^o$ \\
        \midrule
      e  & ... & $0^o$ & $55$ & $1$ & $-21\%$  & $...$ & $15\%$  & $0\%$  
        & $0.08$  & $0.31$ & $0.0^o$ & $0.0^o$ \\
        \midrule
      f  & $180^o$ & $0^o$ & $55$ & $2$ & $-24\%$ & $2\%$ & $8\%$ & 
        $2\%$ & $0.07$ & $0.17$ & $0.0^o$ & $0.0^o$ \\
        \midrule
      g  & $60^o$ & $30^o$ & $55$ & $2$ & $-10\%$ & $-8\%$ & $9\%$ & 
        $11\%$ & $0.06$ & $0.05$ & $9.2^o$ & $58.9^o$ \\
        \midrule
      h  & $-60^o$ & $30^o$ & $55$ & $2$ & $-17\%$ & $2\%$ & $10\%$ & 
        $5\%$ & $0.06$ & $0.06$ & $6.0^o$ & $61.6^o$ \\
      \bottomrule[1.25pt]
    \end{tabular}
    \end{center}
    \smallskip
    \small{Each simulation is labelled by an alphabet letter in the column
      indicated with 'id'. 
      $i_{\rm d2}$ is the mutual inclination between the two disks defined
      by a rotation along the x-axis, $i_{\rm s2}$ is the inclination of
      the hyperbolic trajectory of the second star computed respect to the
      initial plane of the primary disk, $q$ is the minimum approach
      distance during the flyby and $nd$ indicates the number of disks
      included in the simulation.
      At the end of each simulation we have computed the amount of mass
      lost by disk $1$ $\frac{\Delta M_{\rm d1}}{M_{\rm d1}}$ and disk $2$
      $\frac{\Delta M_{\rm d2}}{M_{\rm d2}}$.
      $M_{12}$ and $M_{21}$ are the amount of mass transferred by one disk
      to the other. 
      $e_{\rm d1}$, $e_{\rm d1}$, $i_{\rm d1}$ and $i_{\rm d1}$ are the
      disk eccentricities and inclinations.}
  \end{table*}
    
  \begin{figure}[htbp!]
    \centering
    \includegraphics[width=0.5\textwidth]{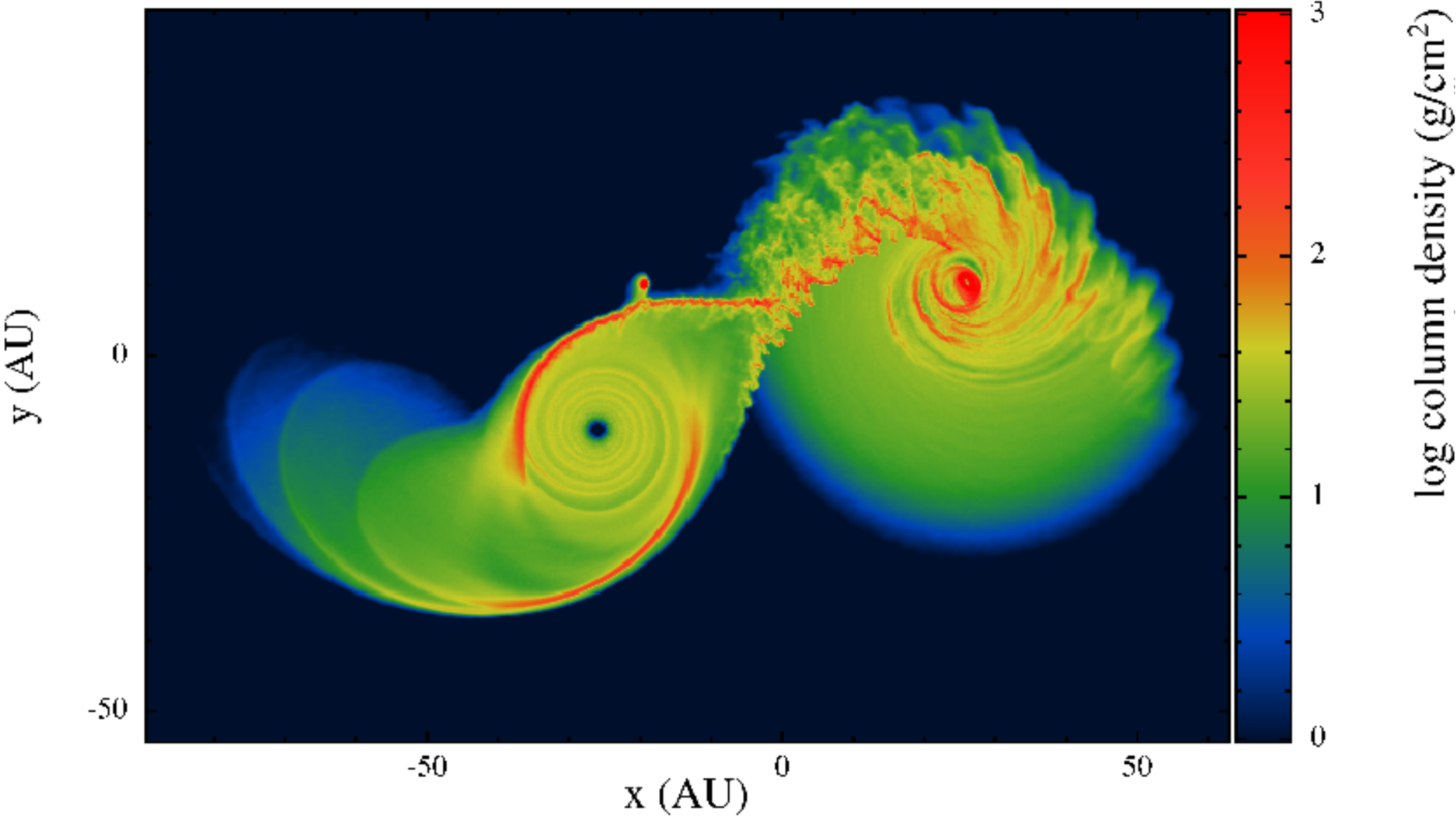}
    \includegraphics[width=0.5\textwidth]{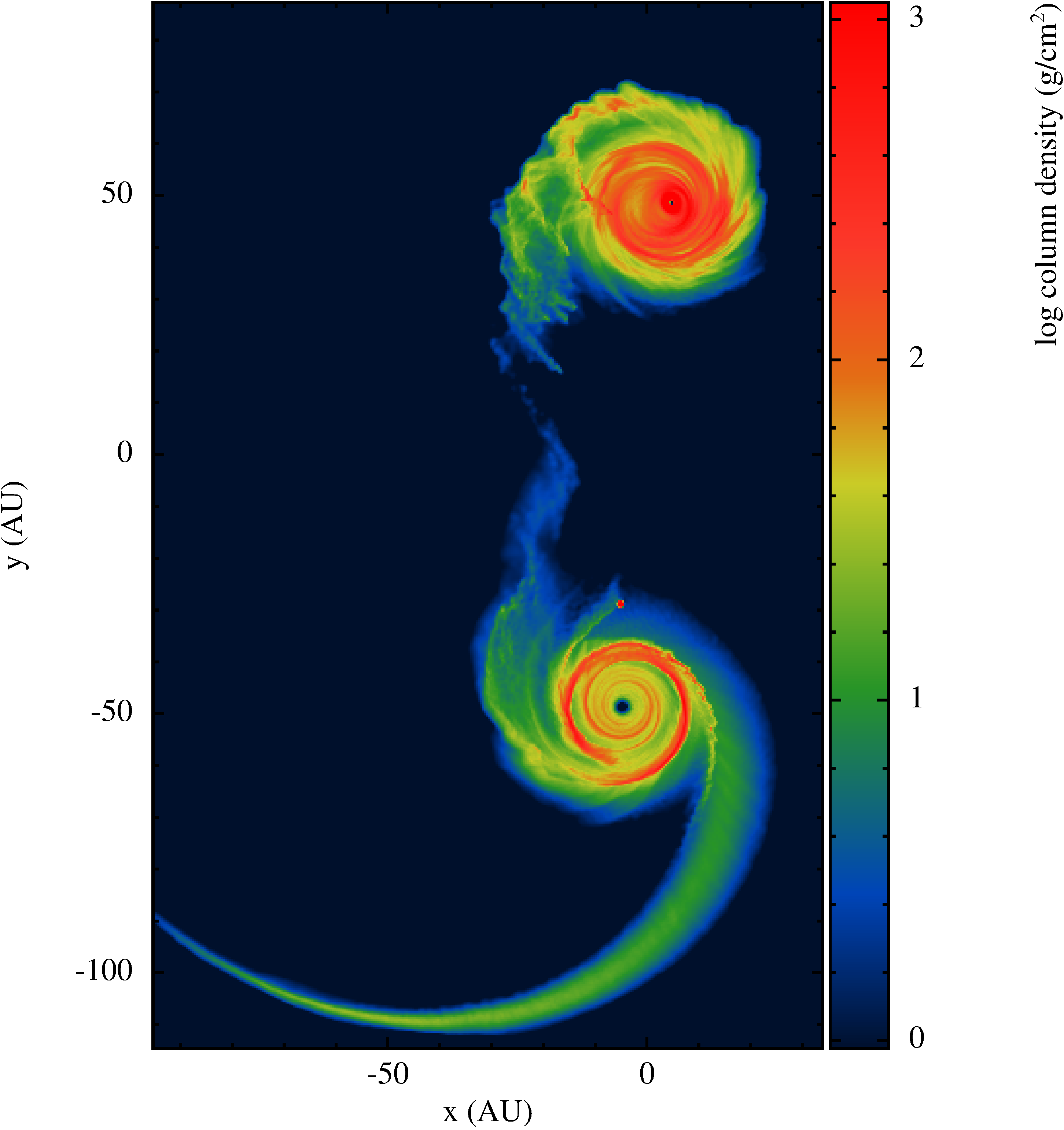}
    \caption{Logarithm of the superficial density (column integrated) of
      the circumstellar disks at $2$ different evolutionary steps  for
      the model e.
      Due to the geometry of the system, the evolution of the two disks is
      asymmetric.}
    \label{fig4}
  \end{figure}

\section{Highly misaligned configurations: cases g and h}

\begin{figure}[htbp!]
    \centering
    \includegraphics[width=0.5\textwidth]{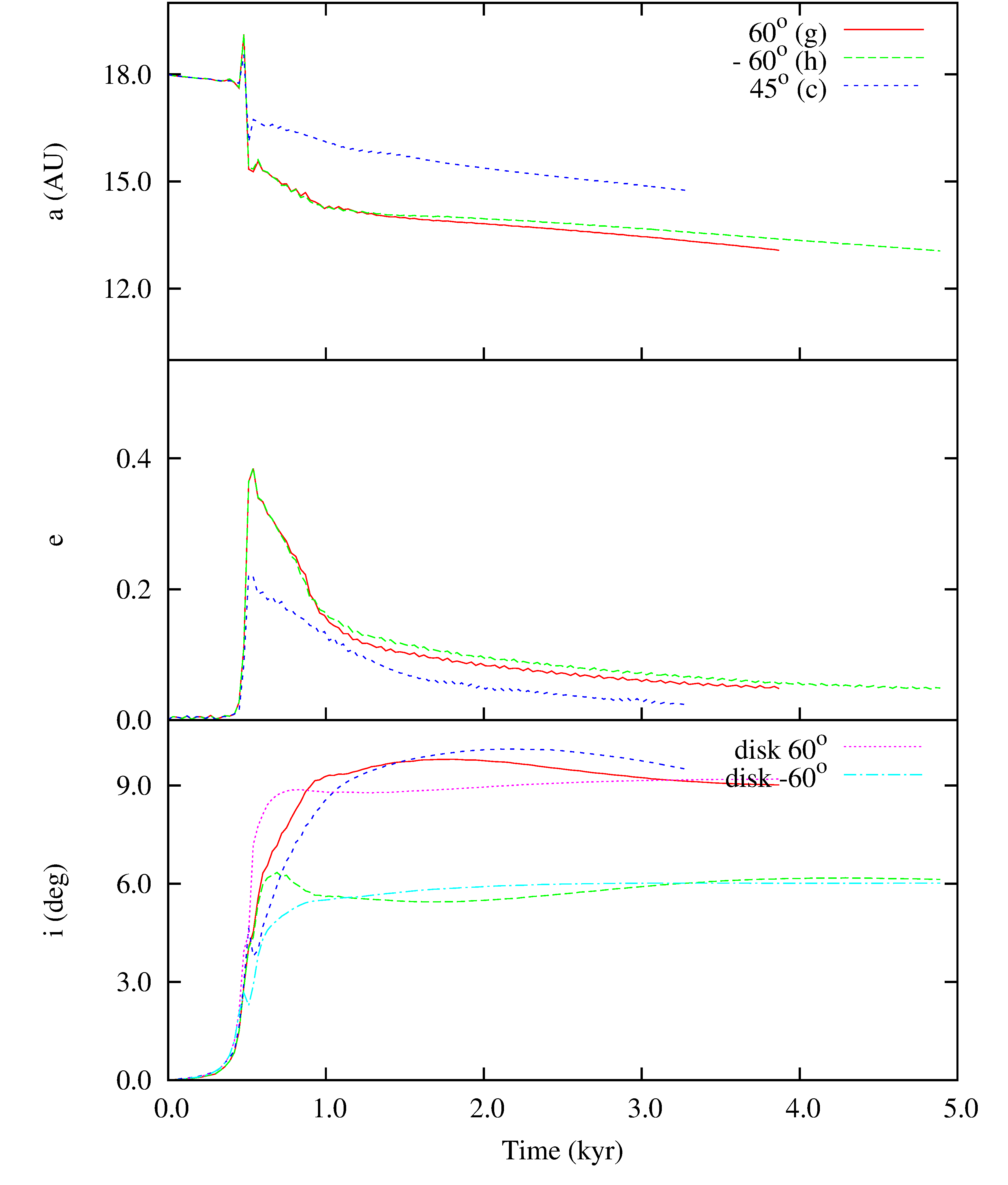}
    \caption{ Evolution of the planet orbital elements ($a$,$e$,$i$)
    in cases g, and h, compared with the test case c.
    In the bottom panel we also plot the inclinations of the primary disks
    in the two models g, and h (labelled disk $60^o$, and disk $-60^o$).
    The bottom panel shows that the planet is dragged to the disk plane on
    a short timescale ($\sim 3$ Kyr).}
    \label{fig5}
  \end{figure}

  We finally considered two cases with a large mutual inclination between
  the disks and an inclined hyperbolic trajectory for the star. 
  In cases g and h $i_{\rm d2}=60^o$ and $i_{\rm d2}=-60^o$, 
  respectively, while $i_{\rm s2} = 30^o$. 
  In Fig.~\ref{fig5} we illustrate the orbital evolution of the planet.
  The evolution of the semimajor axis and eccentricity is very similar to
  that of the other cases previously discussed.
  It is noteworthy that the initial eccentricity step is inversely
  proportional to $i_{\rm s2}$, the inclination of the hyperbolic orbit 
  respect to the primary disk plane. 
  When $i_{\rm s2} = 0^o$ the jump is about $0.5$, for $i_{\rm s2} = 30^o$
  it is slightly smaller than $0.4$ while it reduces to $0.2$ when 
  $i_{\rm s2} = 60^o$.
  A similar trend was observed also by \cite{forgan09} and it may then be
  unnecessary to test higher inclined configurations since the $\Delta e$
  would be smaller. 

  A small difference between cases g and h in terms of eccentricity 
  damping rate can be ascribed to the different mass loss of the two 
  disks.
  In case g the disk mass is reduced by $10\%$ after about $3000$ yr
  while in case h $17\%$ of the mass is lost after the same time
  interval (Tab.~\ref{numb}). 

  As in case c, the mutual inclination between the planet and disk is
  reduced and the planet is quickly pulled back into the disk 
  (Fig.~\ref{fig5} bottom plot).
  Even in this highly inclined configuration, the maximum inclination of
  the disk (and planet) around the primary star is excited to a maximum of
  about $9^o$ (case g) which will be the final misalignment between the
  equatorial plane of the star and the final planet orbit after the disk
  dissipation. 

\section{Conclusions}
\label{results}

  The 3D simulations of stellar encounters in the presence of 
  circumstellar disks, presented in this paper, confirm that the orbital
  changes a planet undergoes during a close stellar flyby are quickly
  damped by the interaction with the disk.
  The only observable difference after the close approach may be a
  residual misalignment between the planet--disk orbital plane and the
  star equatorial plane.
  In fact, during the encounter, the disk harbouring the planet is warped
  by the interaction with the passing star and its disk.
  Gradually, it relaxes on a new plane slightly tilted respect to the original one.
  The amount of tilting depends on the geometrical configuration of the 
  system prior the encounter and in our models we find a maximum value of 
  about $9^o$.
  Subsequently, the planet is pulled back into the disk by the tidal
  interaction with the gas and its orbit also becomes inclined respect to
  the star equatorial plane.
  In case of repeated stellar flybys the final tilting of the planet
  orbit may be larger than  $9^o$  but it appears unlikely that this mechanism may 
  be responsible for 
  the large planet inclinations found in some 
  extrasolar systems \citep{triaud10}. The inclination evolution due to a series of 
  stellar flybys would appear as a random walk 
  with jumps that may lead to positive or negative changes of the planet (and disk) 
  inclination. The stepsize of this random walk would also be small, since 
  we consider an extreme case with a very deep flyby. An event like this 
  would be rare in the lifetime of the star within the cluster and its probability 
  would rapidly decrease with time.

  We find some significant differences between 2D and 3D simulations.
  \begin{itemize}
    \item The mass lost by the primary disk during the stellar flyby  
      depends on the 3D geometry of the encounter and does not lead to the 
      large mass loss observed in 2D simulations.
      The difference is mainly due to the inclusion of a disk around the
      passing star that exchanges mass with the primary disk during the
      close encounter and the use of an SPH code that allows us to better
      track the evolution of the gas far away from the stars. 
    \item The planet migrates inward at a slower rate in 3D simulations
      compared to 2D models.
      The migration of an eccentric orbit, like that of the planet
      just after the encounter, appears to be significantly less efficient
      in 3D models. 
    \item The initial eccentricity excited by the stellar flyby is damped
      more quickly in 3D models and this is partly due to the reduced
      amount of mass lost by the disk. 
  \end{itemize}
  We also observed a slower eccentricity damping efficiency when the orbit
  of the planet is inclined respect to the disk due to a slightly weaker
  disk--planet interaction. 

  Concerning the effects of stellar flybys on planetary systems in stellar
  cluster, we do not expect any signature of the stellar encounters in the
  eccentricity of the planets.
  Some misalignment between the planet orbit and the star equatorial plane
  may appear but it would be limited.
  A period of fast type II migration may occur after the stellar encounter
  since any gap created by the planet before the flyby is destroyed during
  the strong interaction between the stars and the disks at the
  pericenter.
  A new gap is created when the planet is finally pulled back into the
  disk midplane and its orbit circularized. 
  The rate at which the disks are destroyed in clusters may be slower than
  expected since when two stars have an encounter, their disks exchange
  mass rather than loosing it. 

  \cite{marpi13} explored the consequences of stellar encounters on
  systems of 3 planets and a  similar exploration would be interesting also in 3D
  to test the occurrence of a ``jumping Jupiter'' chaotic phase and 
  derive its
  final outcome.
  The parameter space is really huge and many different configurations may
  be envisaged in light of recent discoveries of extrasolar planetary
  systems. 
  However, according to our results, it appears reasonable to expect that
  if the damping works in the extreme conditions we have simulated 
  (low disk density, very deep encounter) it should be effective also
  in different, potentially more complex, scenarios. 

\section*{ACKNOWLEDGMENTS}

\bibliographystyle{aa}
\bibliography{hyper}

\end{document}